\documentclass[aps,12pt,preprintnumbers]{revtex4}
\usepackage{amsmath,amssymb}
\usepackage{bm}
\usepackage{color}
\usepackage[toc,page]{appendix}

\renewcommand\d{\partial}

\begin{document}
\preprint{EFI-14-29}

\title{Hydrodynamics on the lowest Landau level}
\author{Michael Geracie}
\affiliation{Kadanoff Center for Theoretical Physics, University of 
Chicago, Chicago, Illinois 60637, USA}
\author{Dam Thanh Son}
\affiliation{Kadanoff Center for Theoretical Physics, University of 
Chicago, Chicago, Illinois 60637, USA}

\begin{abstract}
Using the recently developed approach to quantum Hall physics
based on Newton-Cartan geometry, we consider the hydrodynamics of an
interacting system on the lowest Landau level.  We rephrase the
non-relativistic fluid equations of motion in a manner that manifests
the spacetime diffeomorphism invariance of the underlying
theory.  In the massless (or lowest Landau level) limit, the fluid
obeys a force-free constraint which fixes the charge current.  An 
entropy current analysis further constrains
the energy response, determining four transverse response functions in
terms of only two: an energy magnetization and a thermal Hall
conductivity. Kubo formulas are presented for all transport
coefficients and constraints from Weyl invariance derived. We also
present a number of St\v reda-type formulas for the equilibrium 
response
to external electric, magnetic and gravitational fields.
\end{abstract}

\maketitle

\section{Introduction}
Since its discovery, the fractional quantum Hall (FQH) 
effect~\cite{Tsui:1982yy,Laughlin:1983} has been subjected to intense 
study and proven a fruitful playground for new concepts in both 
condensed matter and high energy physics. Beyond its quantized Hall 
conductance, FQH states exhibit a number of interesting 
features including anyonic 
excitations~\cite{Halperin:1984fn,Arovas:1984qr}, edge 
states~\cite{Wen:1992vi} and are prime examples of topological phases 
of matter~\cite{Wen-book}.  Beginning with the Laughlin 
wavefunction~\cite{Laughlin:1983}, it has been attacked with numerous 
approaches including Chern-Simons field 
theory~\cite{Zhang:1988wy,Halperin:1992mh} and the composite-fermion 
approach~\cite{Jain:1989}.

This paper follows previous work~\cite{Hoyos:2011ez,Son:2013rqa} with
a focus on spacetime symmetries. This approach has allowed one to
derive a number of new results, including a relationship between the
Hall conductivity at finite wave numbers and the
shift~\cite{Hoyos:2011ez}, sum rules involving the spectral densities
of the stress tensor components~\cite{Golkar:2013gqa}, and a
relationship between the density-curvature response and the chiral
central charge~\cite{Gromov:2014gta}. These and related results have
also been obtained independently using other
methods~\cite{Bradlyn:2012,Can:2014ota,Cho:2014vfl}.

The approach was developed further in Ref.~\cite{Geracie:2014}.  In
the approach we will follow here, a FQH system is put
on a curved spatial manifold whose metric may change with time. The
microscopic theory is found to exhibit a coordinate invariance which
can be interpreted as diffeomorphisms of a geometric structure 
called a Newton-Cartan spacetime.  The FQH system thus is viewed as
living in a 2+1 dimensional Newton-Cartan geometry.  This structure
was initially proposed by Cartan as a geometric description of
Newtonian gravity~\cite{Cartan:1923,Cartan:1924,Kuenzle:1972,Ehlers:1981} and is in fact
the natural coordinate invariant setting for non-relativistic physics
in general.  A Newton-Cartan covariant formulation of non-relativistic
superfluids was developed in Ref.~\cite{Carter:1994}.  Our approach differs
significantly from the latter in particular by the introduction of 
torsion and, in principle, can be applied generally to fluids of any
type, although gapped quantum Hall states will be the focus of our
attention here.

One of the most important features of the FQH problem is the presence
of a large magnetic field separating the Landau levels, reducing the
problem, in its most essential limit, to that of interacting particles
confined to the lowest Landau level (LLL).  The LLL limit can also be
realized by taking the massless limit of a non-relativistic theory.
From the point of view of the symmetries of a Newton-Cartan space,
this massless limit is a regular limit.  This important feature allows
one to directly attack the LLL limit of the FQH problem.
In this paper, we construct the hydrodynamic theory describing a
finite-temperature FQH fluid in the LLL.  (At finite temperature, the
FQH plateaux will be smeared out, but for convenience we
will continue to call any interacting systems of particles confined to
the LLL a ``FQH fluid.'')

Hydrodynamics is an effective theory describing the long distance
physics of a system that is in local thermodynamic equilibrium.  In
the standard hydrodynamic theory we then have as variables a
locally defined temperature $T$ and chemical potential $\mu$, as well
as a fluid velocity $v^i$ that vary slowly in space and time. Their
dynamics is given by conservation laws supplemented by 
constitutive relations expanded to some chosen order in derivatives.
This construction of the hydrodynamic theory is simplest in 
relativistic physics, where covariance is manifest in
the equations of motion
\begin{align}\label{rel eom}
 	\nabla_\mu j^\mu = 0, \qquad \nabla_\nu T^{\mu \nu} = F^{\mu 
\nu} j_\nu .
\end{align}
and simple to implement in the constitutive relations, which are the 
most general
expressions for $j^\mu$ and $T^{\mu\nu}$ in terms of the fluid
degrees of freedom $T$, $\mu$, and $u^\mu$ (the time-like
four-velocity normalized so that $u_\mu u^\mu =1$).  One
merely writes down all possible terms that have the correct index
structure and may be obtained from $T$, $\mu$, $u^\mu$ and $F_{\mu
  \nu}$ only through contraction and differentiation; although the 
second law
of thermodynamics puts extra constraints on this expansion.

In non-relativistic physics, we have three conservation laws: those of
 particle number, momentum and energy (which are independent in this 
context)
\begin{subequations}\label{standard eom}
\begin{align}
	&\partial_0 j^0_\text{nc} + \partial_i j^i_\text{nc} = 0,  \\
	&\partial_0 \varepsilon^0_\text{nc} + \partial_i 
\varepsilon^i_\text{nc} = F_{i0} j^i, \\
	&m \partial_0 j^i_\text{nc} + \partial_j T^{ij}_\text{nc} = 
F^{i0} j_{\text{nc} 0} + F^{ij} j_{\text{nc} j}.
\end{align}
\end{subequations}
We would like the
most general constitutive relations for $j^i_\text{nc}$,
$\varepsilon^i_\text{nc}$, and $T^{ij}_\text{nc}$.  The Galilean
invariance of the equations is then imposed as an additional
constraint.

Newton-Cartan geometry greatly simplifies the process of writing down 
the conservation laws and the constitutive relations. Currents that 
transform 
covariantly under diffeomorphism can be defined and 
covariant Ward identities derived Ref.~\cite{Geracie:2014}. (In contrast, 
the energy current $\varepsilon^\mu_\text{nc}$ and stress 
$T^{ij}_\text{nc}$ are not 
spacetime covariant; the ``nc"'s in 
Eqs.~(\ref{standard eom}) are to distinguish the standard currents 
from the covariant ones we will be using throughout.)
That a Newton-Cartan geometry naturally includes a source for the 
energy current has been noted in Ref.~\cite{Karch:2012} and used to study 
energy transport in a recent paper~\cite{Gromov:2014}.

Our paper is organized as follows. In section~\ref{sec:review} we 
briefly recap the results of Ref.~\cite{Geracie:2014}. 
The derivative expansion and entropy current analysis then proceed 
entirely along the lines of the relativistic case. Section 
\ref{sec:constitutive relations} obtains the most general constitutive 
relations and derives results of the massless limit. We find the FQH 
system is constrained to be force-free, which has powerful 
implications on the dynamics. In particular, all first order charge 
transport is determined by thermodynamics.

Section~\ref{sec:entropy analysis} contains the entropy current 
analysis, completing the program outlined above. We find that what 
are in principle four independent parity odd response coefficients (on the basis 
of symmetries) are determined by only two: and energy magnetization 
$M_E$ and the Righi-Leduc (or 
thermal Hall) coefficient $c_{RL}$. In all, on trivial Newton-Cartan 
backgrounds (i.e., in flat metric and zero field coupled to the energy density)
 we have 
\begin{align}\label{current summary}
	j^0_\text{nc} &= n, \qquad 
	\varepsilon^0_\text{nc} = \epsilon ,\nonumber \\
	j^i_\text{nc} &= \epsilon^{ij} \left( \frac{n}{B} \left(E_j - 
\d_j \mu \right) - \frac{s}{B}\d_j T + \d_j  M \right), \nonumber  \\
	\varepsilon^i_\text{nc} &= 
	\Sigma_T \partial^i  T  + \epsilon^{ij} \left(  \frac{\epsilon 
+ p}{B} \left( E_j - \d_j   \mu  \right) - M \d_j \mu 
	-  T c_{RL} \partial_j T 
	+ \partial_j  M_E  \right), \nonumber \\
	T^{ij}_\text{nc} &= \left( p - \zeta \Theta \right) 
\delta^{ij} - \eta \sigma^{ij} - \tilde \eta \tilde \sigma^{ij} .
\end{align}
Here $T$ is the temperature, $\mu$ the chemical potential and $E_i$, 
$B$ the external electric and magnetic fields. $p$, $n$, $\epsilon$, s 
and $M$ are identified with the internal pressure, number density, 
energy density, entropy density and magnetization density while 
$\zeta$, $\eta$ and $\tilde \eta$ are the usual bulk, shear and Hall 
viscosities and $\Sigma_T$ the thermal conductivity. All are arbitrary 
functions of the thermodynamic variables $T$, $\mu$ and $B$ except for 
constraints from the usual thermodynamic identities and several 
positivity conditions
\begin{align}
	\zeta , \eta \geq 0,  \qquad \qquad
	\Sigma_T \leq 0 .
\end{align} 
The system is dissipationless if and only if all inequalities are 
saturated. Kubo formulas for all coefficients may be found in 
sections~\ref{sec:Kubo} (where they are presented in the Newton-Cartan 
formalism used throughout this paper) and \ref{sec:kubo nc} (where 
they are given in standard form).

A recent analysis of 2+1 dimensional gapped phases derives the most general set of transport coefficients for zero temperature nondissipative systems \cite{Bradlyn:2014}.  Equations (\ref{current summary}) generalize this to to an arbitrary hydrodynamic theory with nonzero temperature and chemical potential (though they are assumed to be slowly varying and far below the gap) giving us the dissipative viscosities and Righi-Leduc coefficient. 

Finally, we present a set of generalized St\v reda formulas that 
characterize the equilibrium response to probing electric, magnetic 
and gravitational fields. A FQH fluid in thermodynamic equilibrium 
has nonzero electric and energy currents,
\begin{align}
	j^i_\text{nc} &=  \varepsilon^{ij} \big( \sigma^{\text{eq}}_H 
E_j  +  \sigma^{B\text{eq}}_H \d_j B + \sigma^{G\text{eq}}_H G_j \big),
\nonumber \\
	\varepsilon^i_\text{nc} &=  \varepsilon^{ij} \big( 
\kappa^{\text{eq}}_H E_j  +  \kappa^{B\text{eq}}_H \d_j B + 
\kappa^{G\text{eq}}_H G_j \big),
\end{align}
where
\begin{align}
	\sigma^\text{eq}_H &= \left( \frac{\d n}{\d B} \right)_{T,\mu},
\qquad \qquad \qquad ~~
	\sigma^{B\text{eq}}_H = \left( \frac{\d M}{\d B} 
\right)_{T,\mu}, \nonumber \\
	\sigma^{G\text{eq}}_H &= T \left( \frac{\d s}{\d B} 
\right)_{T,\mu} + \mu \left( \frac{\d n}{\d B}\right)_{T , \mu} - M,
\nonumber \\
	\kappa^\text{eq}_H &= \left( \frac{\d M_E}{\d \mu} 
\right)_{T,B} - M, \qquad \qquad
	\kappa^{B\text{eq}}_H = \left( \frac{\d M_E}{\d B} 
\right)_{T,\mu}, \nonumber \\
	\kappa^{G\text{eq}}_H &= T\left( \frac{\d M_E}{\d T} 
\right)_{\mu , B}  + \mu \left( \frac{\d M_E}{\d \mu} \right)_{T , B} 
- 2 M_E ,
\end{align}
the first of which may be recognized as the usual St\v reda
formula~\cite{Streda:1982}. Here $G_i = \partial_i \Phi$ is the
external force exerted by a gravitational potential $- \Phi$.

We give concluding remarks in section~\ref{sec:concl}. The appendices 
contain additional constraints due to Weyl invariance and 
other materials of a technical character. 
In a companion paper~\cite{Geracie:noncommutative} we present an 
alternative derivation of some of the results of this paper without 
the use of the Newton-Cartan formalism, compute the thermal Hall coefficient
in the high-temperature regime and discuss the question of 
particle-hole symmetry of the hydrodynamic theory.

\section{Ward Identities}\label{sec:review}

We begin with a brief recap of recent work on the Ward identities of
non-relativistic systems. For details we refer the reader to
Ref.~\cite{Geracie:2014}.  In this paper we derive covariant Ward
identities using the Newton-Cartan structure of non-relativistic
theories~\cite{Cartan:1923,Cartan:1924,Kuenzle:1972,Ehlers:1981}. In 
considering response to a perturbing gravitational scalar potential we 
will need a torsionful version of this geometry (this has also 
been considered in Ref.~\cite{Christensen:2013}).  This
involves a degenerate metric $g^{\mu \nu}$ that measures spatial
distances. It's degeneracy direction is spanned by a one-form $n_\mu$
satisfying $n \wedge dn = 0$ that provides an absolute notion of space 
through it's integral submanifolds. It's convenient to also define an 
auxiliary
``velocity" field $v^\mu$ satisfying $n_\mu v^\mu = 1$ that allows one
to invert the metric to a transverse projector
\begin{align}
	g_{\mu \lambda} g^{\lambda \nu} = {P_\mu}^\nu \qquad 
\text{where} \qquad
	{P_\mu}^\nu = {\delta_\mu}^\nu - n_\mu v^\nu .
\end{align}
 The connection $\nabla_\mu$ is then uniquely specified by
\begin{align}
	\nabla_\mu n_\nu = 0, \qquad
	\nabla_\lambda g^{\mu \nu} = 0, \qquad
	g_{\lambda [ \mu} \nabla_{\nu ]} v^\lambda = 0,
\end{align}
and has torsion ${T^\lambda}_{\mu \nu} = v^\lambda (dn)_{\mu \nu}$. 
The velocity field is unphysical and may be chosen in whatever manner 
is convenient for a particular problem.

In Ref.~\cite{Geracie:2014} we demonstrate that for systems 
constrained to the LLL, the Ward identities following from gauge and 
diffeomorphism invariance in a nonrelativistic theory take the 
covariant form
\begin{align}
	\left( \nabla_\mu - G_\mu \right) j^\mu &= 0, \label{charge 
conservation}\\
	\left( \nabla_\mu - G_\mu \right) \varepsilon^\mu &= - F_{\mu 
\nu} v^\mu j^\nu + G_{\mu \nu} v^\mu \varepsilon^\nu - \frac{1}{2} 
\tau_{\mu \nu} T^{\mu \nu}, \\
	\left( \nabla_\nu - G_\nu \right) T^{\mu \nu} &= {F^\mu}_\nu 
j^\nu - {G^\mu}_\nu \varepsilon^\nu . \label{stress conservation}
\end{align}
Ward identities for Newton-Cartan diffeomorphisms have also been 
considered in Refs.~\cite{Andreev:2013qsa} and \cite{Karch:2012}. The above 
is a covariant generalization of these equations to arbitrary 
backgrounds, subjected to a LLL projection in the form of a massless 
limit. These identities also assume a spinful fluid of spin $s=1$.

Here $j^\mu$ and $\varepsilon^\mu$ are the particle and energy 
currents and $T^{\mu \nu}$ a transverse symmetric stress
\begin{align}
	T^{\mu \nu} n_\nu = 0 .
\end{align}
The stress is conserved except for the action of external forces. The 
first of these is exerted by the familiar electromagnetic field 
strength $F_{\mu \nu} = (dA)_{\mu \nu}$, but there is also a torsional 
field strength $G_{\mu \nu} = ( d n )_{\mu \nu}$ that couples to the 
energy current. Before the LLL projection the equation for stress 
conservation contains terms involving the momentum current. These 
however drop out upon taking the massless limit $m \rightarrow 0$ and 
stress conservation becomes the force balance (\ref{stress 
conservation}).

The first equation expresses conservation of charge current while the 
second is the work-energy equation. Here
\begin{align}\label{shear}
	\tau_{\mu \nu} = \pounds_v g_{\mu \nu} 
\end{align}
is the shear tensor. Although the Ward identities appear to depend on 
a choice of $v^\mu$, one can demonstrate that the implicit and 
explicit dependence cancel and they are in fact invariant under 
$v^\mu$ redefinitions. Finally note in all cases the divergence 
operator takes the form $\nabla_\mu - G_\mu$ where $G_\mu = 
{T^\nu}_{\nu \mu}$ which is the correct form of the divergence on a 
torsionful manifold.

In writing these formulas, we have chosen $g$-factor $g=2$ and spin 
$s=1$ as we are always free to do. The former is necessary for a 
regular massless limit, the later is a matter of convenience. A given 
system may not satisfy these conditions, but in 
Ref.~\cite{Geracie:2014} we present a precise dictionary that allows 
one to translate our results to the general case.

\subsection{Coordinate Expressions}\label{sec:coordinates}
To aid in the interpretation of 
Eqs.~(\ref{charge conservation})---(\ref{stress conservation}) 
and comparison to the usual treatment of 
non-relativistic fluid dynamics, we collect here a number of 
coordinate dependent expressions for the above structure. Because we 
demand $n \wedge dn = 0$, a Newton-Cartan geometry admits a convenient 
set of coordinates called global time coordinates (GTC) in which
\begin{align}
	n_\mu =
	\begin{pmatrix}
		e^{-\Phi}, & 0
	\end{pmatrix}, 
\end{align}
for some scalar potential $\Phi$.
It is instructive to have a few coordinate expressions for the 
structure outlined above in GTC. In these coordinates we may generally 
parameterize the metric and velocity vector as
\begin{align}
	g^{\mu \nu} = 
	\begin{pmatrix}
		0 & 0 \\
		0 & g^{ij}
	\end{pmatrix},
	\qquad
	v^\mu = e^\Phi
	\begin{pmatrix}
		1 \\
		v^i
	\end{pmatrix} .
\end{align}
It's then a matter of calculation to show that
\begin{align}
	&g_{\mu \nu} = 
	\begin{pmatrix}
		v^2 & - v_j \\
		- v_i & g_{ij}
	\end{pmatrix} ,
	&& G^\mu = 
	\begin{pmatrix}
		0 \\
		\partial^i \Phi
	\end{pmatrix} ,
	\nonumber \\
	&\tau^{\mu \nu} = 
	\begin{pmatrix}
		0 & 0 \\
		0 &  e^\Phi \big( \nabla^i v^j + \nabla^j v^i - \dot 
g^{ij} \big)
	\end{pmatrix} ,
	&&\nabla_{\!\mu} v^\mu = e^\Phi \Big( \nabla_{\!i}\, v^i + 
\frac{1}{2} g^{ij} \dot g_{ij} \Big) ,
\end{align}
$\nabla_{\!i}$ being the standard spatial connection.

There is a unique volume element $\varepsilon_{\mu \nu \lambda}$ that 
is compatible with the connection,
\begin{align}
	\nabla_{\!\!\rho}\, \varepsilon_{\mu \nu \lambda} = 0 ,
\end{align}
where we specialize to $2+1$ dimensions from this point forward. If we 
define
\begin{align}
	\varepsilon_{\mu \nu} = \varepsilon_{\mu \nu \lambda} 
v^\lambda,
\end{align}
then $\varepsilon^{\mu \nu}$ plays the role of the spatial volume 
element. Again in GTC we have
\begin{align}
	\varepsilon^{\mu \nu} = 
	\begin{pmatrix}
		0 & 0\\
		0 & \varepsilon^{ij}
	\end{pmatrix} ,
	\qquad
	\varepsilon_{012} = \sqrt{g} e^{-\Phi} ,
\end{align}
where $\varepsilon^{ij}$ is the antisymmetric tensor with 
$\varepsilon^{12} = \frac {1}{\sqrt{g}}$.

In GTC the Ward identities then read
\begin{align}
	&\frac{1}{\sqrt{g}} e^\Phi \partial_0 ( \sqrt{g} e^{-\Phi} 
j^0_\text{nc} ) + e^\Phi \nabla_i ( e^{-\Phi} j^i_\text{nc} ) = 0 ,
\notag \\
	&\frac{1}{\sqrt{g}}\partial_0 \big( \sqrt{g} 
\varepsilon^0_{\text{nc}} \big) + \nabla_i  \varepsilon^i_{\text{nc}} 
= E_i 	j^i_\text{nc} + G_i \varepsilon^i_\text{nc}
	- \frac{1}{2} T^{ij}_{\text{nc}} \dot g_{ij} , \notag \\
	&\nabla_j {T_{\text{nc}i}}^j = j^0_\text{nc} E_i + 
\varepsilon_{ij} j^j_\text{nc} B + \left( \varepsilon^0_{\text{nc}} 
{\delta_i}^j+ {T_i}^j\right) G_j .  
\end{align}
We see that $G_i = \partial_i \Phi$ plays the role of an external 
gravitational field that couples to the energy density so we may think 
of $-\Phi$ as the non-relativistic gravitational potential.

\section{Constitutive Relations}\label{sec:constitutive relations}

In $2+1$ dimensions there are four independent one-point Ward 
identities: current conservation, the work-energy equation and 
Newton's second law. In the low energy, long wavelength limit, we 
expect that the system admits a fluid description, that is, the 
remaining degrees of freedom are also four-fold: two thermodynamic 
variables, which we take to be the temperature $T$ and chemical 
potential $\mu$, and the fluid velocity. The Ward identities then 
suffice to determine the evolution of the system and serve as 
equations 
of motion.

However, in the massless limit we lose two of these degrees of 
freedom. The momentum current drops out of the final Ward identity
\begin{align}
	(\nabla_\nu - G_\nu ) T^{\mu \nu} = {F^\mu}_\nu j^\nu - 
{G^\mu}_\nu \varepsilon^{\nu} , \label{stress}
\end{align}
which now contains no time derivatives. What is typically a dynamical 
equation for the momentum flow reduces to a force-free constraint: 
since the fluid is massless, it is obliged to flow in such a manner 
that the applied forces cancel. We will use this in what follows to 
solve for the charge flow. What remains is two equations of motion
\begin{align}
	(\nabla_\mu - G_\mu ) j^\mu &= 0 , \label{current 
conservation} \\
	(\nabla_\mu - G_\mu ) \varepsilon^\mu = - F_{\mu \nu} v^\mu 
j^\nu &+ G_{\mu \nu} v^\mu j^\nu - \frac12 \tau^{\mu \nu} T_{\mu \nu}, 
\label{work energy} 
\end{align}
that will determine $T$ and $\mu$ for all time given initial 
conditions.

Of course for these to say anything we need to specify constitutive 
relations, that is $T^{\mu \nu}$, $\varepsilon^\mu$ and $j^\mu$ in 
terms of the fluid degrees of freedom $T$, $\mu$, and the external 
fields. In the long wavelength, low energy limit when the fluid 
description is assumed to hold, we can assume that only low powers in 
the derivatives of these variables are important. In this section, we 
present the most general constitutive relations consistent with non-
relativistic diffeomorphism covariance to first order in a derivative 
expansion. 

Our derivative counting scheme for the background fields is as 
follows. The FQH problem assumes a large, nonvanishing magnetic field, 
which we will take to vary slowly in space and time. The fluid is 
also assumed to be moving in a nearly flat geometry and to have only 
slightly departed from thermodynamic equilibrium; that is, $F_{\mu 
\nu}$, $g^{\mu \nu}$, $T$ and $\mu$ are all $\mathcal O (0)$. By it's 
definition $( \nabla_\mu \nabla_\nu - \nabla_\nu \nabla_\mu ) f = - 
{T^\lambda}_{\mu \nu} \nabla_\lambda f$, the torsion must already be 
at $\mathcal O (1)$.

To organize the independent data appearing at each order we first note 
a few convenient facts. To begin, any vector $w^\mu$ may be uniquely 
decomposed into a part parallel to $v^\mu$ and perpendicular to 
$n_\mu$
\begin{align}
	w^\mu = a v^\mu + b^\mu , \qquad \text{where}
	\qquad
	n_\mu b^\mu = 0 .
\end{align} 
A similar decomposition may be carried out for tensors of all types. 
In particular for $(2 , 0)$ tensors we have
\begin{align}
	t^{\mu \nu} = a v^\mu v^\nu + v^\mu b^\mu + c^\mu v^\nu + 
d^{\mu \nu},
\end{align}
for some spatial vectors $b^\mu$ and $c^\mu$ and a spatial tensor 
$d^{\mu \nu}$. As a result, we need only consider scalars, transverse 
vectors, and transverse tensors in our classification. Since 
transverse 
2-tensors may be further decomposed into a trace, a symmetric 
traceless part and an antisymmetric part (which we will not need), we 
are left in the end with scalars, transverse vectors and transverse 
symmetric tensors. We further subdivide this classification by 
evenness or oddness under parity.

\subsection{Zeroth Order}
Let's begin by analyzing the force-free constraint (\ref{stress}). To 
our order we have
\begin{align}
	{F^\mu}_\nu j^\nu = 0 .
\end{align}
The charge current must be proportional to the unique zero eigenvector 
of $F_{\mu \nu}$. We make a ``choice of frame" so that $v^\mu$ tracks 
this equilibrium charge current
\begin{align} \label{equilibrium current 1}
	v^\mu &= \frac{1}{2B} \varepsilon^{\mu \nu \lambda} F_{\nu 
\lambda} 
	=e^\Phi
	\begin{pmatrix}
		1 \\
		\frac{\varepsilon^{ij} E_j}{B} 
	\end{pmatrix} 
	\qquad \implies \qquad
	j^\mu = n v^\mu ,
\end{align}
where $n$ will be some function of the zeroth order data
\begin{center}
\begin{tabular}{ l | c }
		& Data  \\
	Scalar & $T \qquad \mu$   \\
 	Pseudoscalar & $B$ 
\end{tabular}
\end{center}
Here $B=\frac{1}{2} \varepsilon^{\mu \nu} F_{\mu \nu}$ is of course 
the magnetic field.
Note that in this frame we have
\begin{align}
	F_{\mu \nu} = B \varepsilon_{\mu \nu}\,, \qquad \nabla_\mu ( B 
v^\mu ) = 0.
\end{align}

In the reference frame comoving with the charge, we expect that the 
equilibrium state is invariant under spatial rotations. This implies 
that the energy current must coincide with the charge current and the 
stress must be pure trace. Hence
\begin{align}\label{equilibrium current 2}
	\varepsilon^\mu = \epsilon v^\mu , \qquad
	T^{\mu \nu} = p g^{\mu \nu} ,
\end{align}
where $\epsilon$ and $p$ are again functions of $T$, $\mu$ and 
$B$.

However, $p$, $\epsilon$ and $n$ are not entirely arbitrary being
constrained by thermodynamics. It can be shown from statistical
considerations (see appendix \ref{append:mag}) that $\epsilon$ is the
energy density and $n$ number density of the fluid.  The hydrodynamic
pressure $p$ is sometimes
called ``internal pressure" and is related to the grand potential
density (sometimes called the ``thermodynamic 
pressure'') $p_\text{thm} = p_\text{thm} ( T , \mu , B )$ by a Legendre
transformation
\begin{align}
	p = p_\text{thm} - B \partial_B p_\text{thm} .
\end{align}
To simplify some of our formulae we prefer to work with the internal 
pressure $p = p ( T , \mu , M )$ which is naturally a function of $T$, 
$\mu$ and the magnetization density $M = \partial_B p_\text{thm}$. We 
will thus exchange $B$ for $M$ as the independent variable in what 
follows. The functions $p$, $\epsilon$ and $n$ satisfy the 
thermodynamic identities
\begin{align}\label{thermodynamics}
	\epsilon + p = T s + \mu n - M B,
	\qquad
	d \epsilon = T ds + \mu dn - M dB .
\end{align}
Only one of these functions (say $p$) is independent. It is called the 
equation of state.

\subsection{First Order}\label{sec:constraints}
We now seek the most general corrections to $T^{\mu \nu}$, 
$\varepsilon^\mu$ and $j^\mu$ to first order. Denote these as
\begin{align}
	j^\mu = n v^\mu + \nu^\mu , \qquad
	\varepsilon^\mu = \epsilon v^\mu + \xi^\mu , \qquad
	T^{\mu \nu} = p g^{\mu \nu} + \pi^{\mu \nu}.
\end{align}
The complete set of first order data is 
\begin{center}
	\begin{tabular}{ l | c c c}
			& Independent Data  \\
		Scalar & $\Theta \qquad  ( v^\mu \nabla_\mu  T ) 
\qquad ( v^\mu \nabla_\mu \mu )$   \\
 		Pseudoscalar & --  \\
		Vector & $\nabla^\mu T \qquad \nabla^\mu \mu \qquad 
\varepsilon^{\mu \nu} \nabla_\nu M \qquad G^\mu$\\
		Pseudovector & $\varepsilon^{\mu \nu} \nabla_\nu T 
\qquad \varepsilon^{\mu \nu} \nabla_\nu \mu \qquad \nabla^\mu M \qquad 
\varepsilon^{\mu \nu}G_\nu$  \\
		Traceless Symmetric Tensor & $\sigma^{\mu \nu}$  \\
		Traceless Symmetric Pseudotensor & $\tilde \sigma^{\mu 
\nu}$
	\end{tabular}
\end{center}
Here $\Theta = \nabla_\mu v^\mu$ is the expansion and $\sigma^{\mu 
\nu} = \tau^{\mu \nu} - \Theta g^{\mu \nu}$ the traceless shear. 
The ``tilde" operation is defined for symmetric two tensors as $\tilde 
A^{\mu \nu} =\frac{1}{2} \big( A^{\mu \lambda} 
{\varepsilon_\lambda}^\nu+ A^{\nu \lambda}{\varepsilon_\lambda}^\mu 
\big)$. We do not include the material derivative of all three 
thermodynamic variables since one may always be eliminated by the 
constraint
\begin{align}
	\nabla_\mu ( B v^\mu ) = 0 \qquad \implies \qquad
	v^\mu \nabla_\mu B = - B \Theta .
\end{align}

Not all of this data is independent on-shell and we may choose to 
eliminate some in favor of the others by solving the equations of 
motion. In our case, there are two scalar equations: the continuity 
equation and the work-energy equation. We use these to eliminate the 
material derivatives of $T$ and $\mu$, as indicated by parentheses.

Before we continue, a few comments on fluid frames are due. Since we
will be considering small departures from thermal equilibrium there is
an inherent ambiguity at first order in derivatives in how we define
$T$ and $\mu$. This is a problem extensively discussed in the
literature on nonequilibrium
fluids~\cite{Jensen:2012,Bhattacharya:2011}. We differ from the usual
case only in that we do not have any independent definition of a fluid
velocity that would require additional fixing. Hence we have a two
parameter ambiguity which we choose to fix by going to the Landau
frame
\begin{align}
	n_\mu \nu^\mu = n_\mu \xi^\mu = 0 .
\end{align}
Note that we have $\pi^{\mu \nu} n_\nu = 0$ for free since the stress 
is a transverse tensor. 

The most general first order constitutive relations are then
\begin{subequations}\label{constitutive relations}
\begin{align}
	\nu^\mu &=  \chi_T \nabla^\mu T + \chi_{\mu} \nabla^\mu \mu + 
\tilde \chi_M \nabla^\mu M + \chi_G G^\mu \nonumber \\ 
	&\qquad+ \tilde \chi_T \varepsilon^{\mu \nu} \nabla_\nu T + 
\tilde \chi_{\mu} \varepsilon^{\mu \nu} \nabla_\nu \mu + \chi_M 
\varepsilon^{\mu \nu} \nabla_\nu M + \tilde \chi_G \varepsilon^{\mu 
\nu} G_\nu, \\
	\xi^\mu &= \Sigma_T \nabla^\mu T + \Sigma_{\mu} \nabla^\mu \mu 
+ \tilde \Sigma_M \nabla^\mu M + \Sigma_G G^\mu \nonumber \\ 
	&\qquad+ \tilde \Sigma_T \varepsilon^{\mu \nu} \nabla_\nu T + 
\tilde \Sigma_{\mu} \varepsilon^{\mu \nu} \nabla_\nu \mu + \Sigma_M 
\varepsilon^{\mu \nu} \nabla_\nu M + \tilde \Sigma_G  \varepsilon^{\mu 
\nu} G_\nu, \\
	\pi^{\mu \nu} &=  - \zeta \Theta g^{\mu \nu} - \eta 
\sigma^{\mu \nu} - \tilde \eta \tilde \sigma^{\mu \nu} ,
\end{align}
\end{subequations}
where a tilde denotes oddness under parity. We derive Kubo formulas 
for these coefficients in section \ref{sec:Kubo}.

\subsection{Force-Free Flows}
As mentioned previously, we may use the force balance constraint to 
completely solve for the charge current
\begin{align}\label{restrictions1}
	& \qquad \qquad  \nabla^\mu p = B \varepsilon^{\mu \nu} 
\nu_\nu 
+ ( \epsilon + p ) G^\mu \qquad \implies \nonumber \\
	\tilde \chi_T &= - \frac{s}{B}\,, \qquad
	\tilde \chi_\mu = - \frac{n}{B}\,,  \qquad
	\chi_M = 1, \qquad
	\tilde \chi_G =  \frac{\epsilon + p}{B}\, .
\end{align}
All charge transport coefficients are thus determined by the equation 
of 
state. Also note that all longitudinal responses are zero. This is 
because the Lorentz force must cancel forces from pressure gradients 
and 
the magnetic field always produces a force perpendicular to the 
current; 
hence the current must be perpendicular to pressure gradients.

\section{Entropy Current Analysis}\label{sec:entropy analysis}
The constitutive relations (\ref{constitutive relations}) subject to
the restrictions (\ref{restrictions1}) are the most general possible
that are consistent with the equations of motion and
constraint. However, it is still possible to generate flows that
violate the second law of thermodynamics. For example, it is well
known that a negative shear viscosity allows one to remove entropy
from an isolated system and so we should have $\eta \geq 0$
\cite{Landau:1987}. To derive all such restrictions, we perform an
entropy current analysis along the lines of
Ref.~\cite{Jensen:2012}. Lacking a spacetime picture of
non-relativistic physics, previous analyses were restricted to the
Lorentzian case and in particular did not include an independent
energy current. Our results reproduce theirs for those coefficients
that we have in common as well as derive new results for energy
transport.

The canonical entropy current is
\begin{align}
	s^\mu_\text{can} = s v^\mu - \frac{\mu}T\nu^\mu + \frac1T 
\xi^\mu,
\end{align}
but out of equilibrium we should in principle once again expand in 
first order data
\begin{align}
	s^\mu = s^\mu_\text{can} + \zeta^\mu ,
\end{align}
where
\begin{align}
	\zeta^\mu &= \zeta_\Theta \Theta v^\mu + \zeta_T \nabla^\mu T 
+ \zeta_{\mu} \nabla^\mu \mu + \tilde \zeta_M \nabla^\mu M + \zeta_G 
G^\mu \nonumber \\ 
	&+ \tilde \zeta_T \varepsilon^{\mu \nu} \nabla_\nu T + \tilde 
\zeta_{\mu} \varepsilon^{\mu \nu} \nabla_\nu \mu + \zeta_M 
\varepsilon^{\mu \nu} \nabla_\nu M + \tilde \zeta_G \varepsilon^{\mu 
\nu} G_\nu .
\end{align}
Now we impose the second law. For non-negative entropy production 
between all spatial slices, we must have
\begin{align}\label{second law}
	(\nabla_\mu - G_\mu ) s^\mu \geq 0 .
\end{align}

Using the equations of motion in the form
\begin{subequations}
\begin{align}
	&v^\mu \nabla_\mu n + n \Theta = -\nabla_\mu \nu^\mu + G_\mu 
\nu^\mu, \\
	&v^\mu \nabla_\mu \epsilon + ( \epsilon + p ) \Theta = - 
\nabla_\mu \xi^\mu - \frac{1}{2} \Theta \pi - \frac{1}{2} \sigma^{\mu 
\nu} \pi_{\mu \nu} + 2 G_\mu \xi^\mu ,
\end{align}
\end{subequations}
one may check that the divergence of the canonical entropy current is 
a quadratic form in first order data
\begin{align}
	( \nabla_\mu - G_\mu )  s^\mu_{\text{can}} = - \nu^\mu 
\nabla_\mu \Big( \frac{\mu}{T} \Big)- \frac{1}{2T} \Theta g^{\mu \nu} 
\pi_{\mu \nu} - 
\frac{1}{2T} \sigma^{\mu \nu} \pi_{\mu \nu} - \frac{1}{T^2} \xi^\mu ( 
\nabla_\mu T - T G_\mu ) ,   
\end{align}
and so the only genuine second order data in (\ref{second law}) is
\begin{align}
	\nabla_\mu \zeta^\mu \Big|_{2 - \partial} = \zeta_\Theta v^\mu 
\nabla_\mu \Theta  + \zeta_T \nabla^2 T + \zeta_{\mu} \nabla^2 \mu + 
\tilde \zeta_M \nabla^2 M + \zeta_G  \nabla_\mu G^\mu ,
\end{align}
where we have used the Newton-Cartan identities $\varepsilon^{\mu \nu} 
G_{\mu \nu} = 0$ and $\varepsilon^{\mu \nu} \nabla_\mu G_\nu = 0$. 
Since each term may be independently varied to have either sign, all 
coefficients appearing in this equation must be zero.

The remaining first order data is then
\begin{align}\label{entropy current analysis}
	(\nabla_\mu - G_\mu ) s^\mu 
	&= \frac{1}{T} \zeta \Theta^2 + \frac{1}{2T} \eta \sigma_{\mu 
\nu} \sigma^{\mu \nu} + \frac{1}{T} \Sigma_G G_\mu G^\mu \nonumber \\
	&+ \frac{1}{T} \Big( \Sigma_T - \frac{1}{T} \Sigma_G \Big) 
G^\mu \nabla_\mu T + \frac{1}{T} \Sigma_\mu G^\mu \nabla_\mu \mu + 
\frac{1}{T} \tilde \Sigma_M G^\mu \nabla_\mu M \nonumber \\
	&- \Big( \partial_T \tilde \zeta_G + \tilde \zeta_T - \frac{1}
{T}\tilde \Sigma_T - \frac{1}{T^2} \tilde \Sigma_G + \frac{\mu}{T^2} 
\tilde \chi_G \Big)  \varepsilon^{\mu \nu} G_\mu \nabla_\nu  T 
\nonumber \\
	&- \Big( \partial_\mu \tilde \zeta_G + \tilde \zeta_\mu - 
\frac{1}{T}\tilde \Sigma_\mu - \frac{1}{T} \tilde \chi_G \Big) 
\varepsilon^{\mu \nu} G_\mu \nabla_\nu \mu
	- \Big( \partial_M \tilde \zeta_G + \zeta_M - \frac{1}{T} 
\Sigma_M \Big)  \varepsilon^{\mu \nu} G_\mu \nabla_\nu  M \nonumber \\
	&- \frac{1}{T^2} \Sigma_T \nabla_\mu T \nabla^\mu T - \frac{1}
{T^2} \Sigma_\mu \nabla_\mu T \nabla^\mu \mu - \frac{1}{T^2} \tilde 
\Sigma_M \nabla_\mu T \nabla^\mu M \nonumber \\
	& +\Big( \partial_T \tilde \zeta_\mu - \partial_\mu \tilde 
\zeta_T + \frac{1}{T} \tilde \chi_T + \frac{\mu}{T^2} \tilde \chi_\mu 
- \frac{1}{T^2} \tilde \Sigma_\mu \Big) \varepsilon^{\mu \nu} 
\nabla_\mu T \nabla_\nu \mu \nonumber \\
	& +\Big( \partial_T \zeta_M - \partial_M \tilde \zeta_T + 
\frac{\mu}{T^2} \chi_M - \frac{1}{T^2} \Sigma_M \Big) \varepsilon^{\mu 
\nu} \nabla_\mu T \nabla_\nu M \nonumber \\
	&+\Big( \partial_\mu \zeta_M - \partial_M \tilde \zeta_\mu - 
\frac{1}{T} \chi_M \Big) \varepsilon^{\mu \nu} \nabla_\mu \mu 
\nabla_\nu M \nonumber \\
	&\geq 0.
\end{align}
Note that by $\partial_\mu$ we mean the partial derivative with 
respect to the chemical potential, not a spatial derivative. For 
clarity we will always use $\nabla_\mu$ for the spatial derivative 
when there is the possibility of confusion.

The $\nabla_\mu T \nabla^\mu T$, $G^\mu \nabla_\mu T$ and $G_\mu 
G^\mu$ terms need not be separately constrained. We obtain a less 
stringent condition by setting $\Sigma_T = - \frac{1}{T} \Sigma_G$, in 
which case they arrange into a perfect square
\begin{align}
	- \frac{1}{T^2} \Sigma_T (\nabla_\mu T - T G_\mu) (\nabla^\mu 
T - T G^\mu) .
\end{align}
We note in passing that in thermal equilibrium there can be no entropy 
production. This implies
\begin{align}\label{equilibrium T}
	\nabla^\mu T = T G^\mu ,
\end{align}
or $\partial_i T = T \partial_i \Phi$ in coordinates. The physics of 
this clear: $-\Phi$ is the source that couples to the energy density 
and plays the role of a Newtonian gravitational potential. Heat will 
tend to flow from regions of higher $-\Phi$ to lower $-\Phi$. 
Equilibrium is reached once the temperature profile is such that 
(\ref{equilibrium T}) is satisfied. This result is also follows from 
the treatment of equilibrium statistical mechanics in appendix 
\ref{append:mag}. In general relativity this is known as the Tolman-
Ehrenfest effect which states that the redshifted temperature $T || 
\xi 
||$ is constant in thermal equilibrium for $\xi$ a timelike killing 
field \cite{Tolman:1930}. In the non-relativistic case we have $T 
n_\mu \xi^\mu = \text{const.}$

From (\ref{entropy current analysis}) we immediately obtain the 
expected signs of the parity even viscosities and thermal conductivity
\begin{align}
	\zeta \geq 0, \qquad
	\eta \geq 0, \qquad
	\Sigma_T \leq 0 .
\end{align}
The remaining terms place new restrictions on the energy and entropy 
coefficients
\begin{align}\label{restrictions}
	\Sigma_\mu = \tilde \Sigma_M = 0, \qquad&
	\begin{pmatrix}
		 \partial_\mu \zeta_M - \partial_ M \tilde \zeta_\mu 
\\
		\partial_M \tilde \zeta_T - \partial_T \zeta_M \\
		\partial_T \tilde \zeta_\mu - \partial_\mu \tilde 
\zeta_T
	\end{pmatrix}
	=
	\begin{pmatrix}
		\frac{1}{T} \chi_M  \\
		- \frac{1}{T^2} \Sigma_M + \frac{\mu}{T^2} \chi_M \\
		\frac{1}{T^2} \tilde \Sigma_\mu - \frac{1}{T} \tilde 
\chi_T - \frac{\mu}{T^2} \tilde \chi_\mu
	\end{pmatrix},
	\nonumber \\
	\begin{pmatrix}
		\partial_T \tilde \zeta_G \\
		\partial_\mu \tilde \zeta_G \\
		\partial_M \tilde \zeta_G
	\end{pmatrix}
	&=
	\begin{pmatrix}
		- \tilde \zeta_T + \frac{1}{T} \tilde \Sigma_T + 
\frac{1}{T^2} \tilde \Sigma_G - \frac{\mu}{T^2} \tilde \chi_G \\
		- \tilde \zeta_\mu + \frac{1}{T} \tilde \Sigma_\mu + 
\frac{1}{T} \tilde \chi_G \\
		- \zeta_M + \frac{1}{T} \Sigma_M
	\end{pmatrix} .
\end{align}

We seek the most general solution to these constraints. Begin by 
eliminating the entropy coefficients by taking the curl of the third 
equation and plugging in the second
\begin{align}\label{PDE}
	\begin{pmatrix}
		 \partial_{\mu} \Big( \frac{1}{T} \Sigma_M \Big) - 
\partial_M \Big( \frac{1}{T}\tilde \Sigma_{ \mu} \Big) \\
		 \partial_M \Big( \frac{1}{T} \tilde \Sigma_T \Big) - 
\partial_T \Big( \frac{1}{T} \Sigma_M \Big) \\
		\partial_T \Big( \frac{1}{T} \tilde \Sigma_{ \mu} 
\Big) - \partial_{ \mu} \Big( \frac{1}{T} \tilde \Sigma_T \Big)
	\end{pmatrix} 
	=
	\begin{pmatrix}
		\frac{1}{T} \big( \chi_M + \partial_M \tilde \chi_G 
\big) \\
		- \frac{1}{T^2} \big( \Sigma_M + \partial_M \tilde 
\Sigma_G ) + \frac{\mu}{T^2} \big( \chi_M + \partial_M \tilde \chi_G 
\big) \\
		\frac{1}{T^2} \big( \tilde \Sigma_\mu + \partial_\mu 
\tilde \Sigma_G \big) - \frac{1}{T} \big( \tilde \chi_T + \partial_T 
\tilde \chi_G \big) - \frac{\mu}{T^2} \big( \tilde \chi_\mu + 
\partial_\mu \tilde \chi_G \big) 
	\end{pmatrix} .
\end{align}
Since the left hand side is the curl of a vector, the right hand side 
is divergenceless and it appears as if we might obtain another 
constraint. However one may check that this is automatically satisfied 
by virtue of the constraints (\ref{restrictions1}) and the 
thermodynamic identities (\ref{thermodynamics}).

We may simplify the partial differential equation (\ref{PDE}) by a 
substitution that isolates the energy response's dependence on the 
equation of state and $\tilde \Sigma_G$
\begin{align}
	&\tilde \Sigma_T = - \frac{1}{T} \tilde \Sigma_G + \frac{\mu}
{T} \frac{Ts + \mu n}{B} + T^2 \tilde g_T, \qquad
	\tilde \Sigma_\mu = - \frac{Ts + \mu n}{B} + T^2 \tilde g_\mu 
, \qquad
	\Sigma_M = T^2 g_M, \nonumber \\
	&\qquad\qquad\qquad \implies \qquad
	\begin{pmatrix}
		\partial_\mu g_M - \partial_M \tilde g_\mu \\
		\partial_M \tilde g_T - \partial_T g_M \\
		\partial_T \tilde g_\mu - \partial_\mu \tilde g_T
	\end{pmatrix}
	=
	\begin{pmatrix}
		0 \\
		0 \\
		0
	\end{pmatrix} .
\end{align}
We see that since $( \tilde g_T , \tilde g_\mu , g_M )$ is curl free, 
it must be 
the gradient of some function
\begin{align}
	\tilde g_T = \partial_T \tilde g ,\qquad
	\tilde g_\mu = \partial_\mu \tilde g , \qquad
	g_M = \partial_M \tilde g .
\end{align}

\subsection*{Summary}\label{sec:summary}
This completes the entropy current analysis. For convenience, we 
collect our results in this section. FQH fluids may be generally 
viewed as massless fluids in a Newton-Cartan geometry. For the special 
values $g=2$, $s=1$ of the parity breaking parameters we have the 
following constitutive relations:

The charge-current response is purely transverse
\begin{align}\label{charge current-summary}
	j^\mu = n v^\mu + \tilde \chi_T \varepsilon^{\mu \nu} 
\nabla_\nu T + \tilde \chi_\mu \varepsilon^{\mu \nu} \nabla_\nu \mu + 
\chi_M \varepsilon^{\mu \nu} \nabla_\nu M + \tilde \chi_G 
\varepsilon^{\mu \nu} G_\nu ,
\end{align}
where all coefficients are determined in terms of thermodynamics
\begin{align}
	\tilde \chi_T = - \frac{s}{B}\,, \qquad
	\tilde \chi_\mu = -\frac{n}{B}\,, \qquad
	\chi_M = 1, \qquad
	\tilde \chi_G = \frac{\epsilon + p}{B} .
\end{align}
Since $v^\mu = e^\Phi ( 1 , \frac{\varepsilon^{ij} E_j}{B})$, we have 
a pure Hall conductivity $\sigma_H = e^\Phi \frac{n}{B}$.

The energy-current takes the form
\begin{align}\label{energy current}
	\varepsilon^\mu = \epsilon v^\mu + \Sigma_T \big( \nabla^\mu T 
- T G^\mu \big) + \tilde \Sigma_T \varepsilon^{\mu \nu} \nabla_\nu T + 
\tilde \Sigma_\mu \varepsilon^{\mu \nu} \nabla_\nu \mu + \Sigma_M 
\varepsilon^{\mu \nu} \nabla_\nu M + \tilde \Sigma_G \varepsilon^{\mu 
\nu} G_\nu .
\end{align}
There is one longitudinal response, the thermal conductivity
\begin{align}
	\Sigma_T \leq 0.
\end{align}
The remaining four coefficients are all transverse and depend only on 
the equation of state and two arbitrary functions $\tilde \Sigma_G$ 
and $\tilde g$ of $T$, $\mu$ and $M$
\begin{subequations}\label{energy coefficients}
\begin{align}
	&\tilde \Sigma_T = - \frac{1}{T} \tilde \Sigma_G + T^2 
\partial_T \tilde g + \frac{\mu}{T} \frac{Ts + \mu n}{B} \,, \\
	&\tilde \Sigma_\mu = T^2 \partial_\mu \tilde g - \frac{Ts+\mu 
n}{B} 
   \,,\qquad
	\Sigma_M = T^2 \partial_M \tilde g.
\end{align}
\end{subequations}
Using (\ref{restrictions}) we find the entropy current is determined 
by $\tilde g$ and $\tilde \zeta_G$
\begin{align}
	s^\mu &= s^\mu_\text{can}
	+ \tilde \zeta_T \varepsilon^{\mu \nu} \nabla_\nu T + \tilde 
\zeta_{\mu} \varepsilon^{\mu \nu} \nabla_\nu \mu + \zeta_M 
\varepsilon^{\mu \nu} \nabla_\nu M + \tilde \zeta_G \varepsilon^{\mu 
\nu} G_\nu ,
\end{align}
where
\begin{subequations}
\begin{align}
	\tilde \zeta_T &= T \partial_T \tilde g - \partial_T \tilde 
\zeta_G + \frac{\mu M}{T^2} \,, \\
	\tilde \zeta_\mu &= T \partial_\mu \tilde g - \partial_\mu 
\tilde \zeta_G - \frac{M}{T} \,, \\
	\zeta_M &= T \partial_M \tilde g - \partial_M\tilde \zeta_G .
\end{align}
\end{subequations}

Finally, the stress is determined by the internal pressure and three 
viscosities
\begin{align}
	T^{\mu \nu} = p g^{\mu \nu} - \zeta \Theta g^{\mu \nu} - \eta 
\sigma^{\mu \nu} - \tilde \eta \tilde \sigma^{\mu \nu} .
\end{align}
The bulk and shear viscosities must be non-negative
\begin{align}
	\zeta \geq 0, \qquad \eta \geq 0 ,
\end{align}
whereas the Hall viscosity $\tilde \eta$ is unconstrained. In a Weyl 
invariant theory, the bulk viscosity must vanish. The complete set of 
restrictions imposed by Weyl on the coefficients considered above 
are given in appendix \ref{append:weyl}.

\section{Kubo Formulas}\label{sec:Kubo}

Fractional quantum Hall transport is determined by $p$, $\zeta$, 
$\eta$, $\tilde \eta$, $\Sigma_T$, $\tilde \Sigma_G$ and $\tilde g$, 
some of which are subject to positivity constraints, but are otherwise 
arbitrary functions of $T$, $\mu$ and $M$. In this section we provide 
Kubo formula's for these functions. For concreteness, perturb around a 
flat background
\begin{align}
	n_\mu = \begin{pmatrix}
		1 & 0 
	\end{pmatrix} ,
	\qquad
	g^{\mu \nu} = 
	\begin{pmatrix}
		0 & 0 \\
		0 & \delta^{ij}
	\end{pmatrix} ,
	\qquad
	F_{\mu \nu} = 
	\begin{pmatrix}
		0 & 0 \\
		0 & \varepsilon_{ij} B
	\end{pmatrix} ,
	\qquad
	v^\mu = 
	\begin{pmatrix}
		1 \\
		0
	\end{pmatrix} .
\end{align}
with wavevector $k_\mu$. The wavevector may be decomposed into 
temporal and transverse parts $k_\mu = \omega n_\mu + q_\mu$. We will 
be considering response in both the ``rapid" ($q^\mu \rightarrow 0$) 
and ``slow" ($\omega \rightarrow 0$) cases.
The two-point functions of interest are
\begin{equation}
	G^{\mu , \nu}_{\varepsilon \varepsilon}  (x) = \frac{\delta 
\left \langle \varepsilon^\mu (x)  \right \rangle}{\delta n_\nu (0)} 
\,, \qquad
	G^{\mu , \nu}_{\varepsilon j}  (x) = \frac{\delta \left 
\langle \varepsilon^\mu  (x) \right \rangle}{\delta A_\nu (0)} \,, 
\qquad
	G^{\mu \nu , \lambda \rho}  (x) = \frac{\delta \left \langle 
T^{\mu \nu}  (x) \right \rangle}{\delta h_{\lambda \rho}(0)} \,.
\end{equation}
Explicitly, these are
\begin{subequations}
\begin{align}
	G^{\mu , \nu}_{\varepsilon \varepsilon} (x) &= \left \langle 
\frac{\delta \varepsilon^\mu (x)}{ \delta n_\nu ( 0 )} \right \rangle 
- i \theta ( x^0 ) \left \langle [ \varepsilon^\mu ( x) , 
\varepsilon^\nu ( 0 )] \right \rangle , \\
	G^{\mu , \nu}_{\varepsilon j} (x) &= \left \langle 
\frac{\delta \varepsilon^\mu (x)}{ \delta A_\nu ( 0 )} \right \rangle 
+ i \theta ( x^0 ) \left \langle [ \varepsilon^\mu ( x) , j^\nu ( 0 )] 
\right \rangle , \\
	G^{\mu \nu , \lambda \rho}(x) &= \left \langle \frac{\delta 
T^{\mu \nu} (x)}{ \delta g_{\lambda \rho} ( 0 )} \right \rangle + 
\frac{i}{2} \theta ( x^0 ) \left \langle [ T^{\mu \nu} ( x) , 
T^{\lambda \rho} ( 0 )] \right \rangle  .
\end{align}
\end{subequations}
The contact terms in these equations do not contribute to the imaginary
parts of the respective Green's functions in momentum space, which 
will appear later in Kubo's formulas.

In this section we prefer to take all coefficients as functions 
of $T$, $\mu$ and $B$ rather than $T$, $\mu$ and $M$ as it is less 
awkward to deal with electromagnetic perturbations. It's a 
straightforward matter to translate the $T$, $\mu$, $M$ dependence of 
equations (\ref{charge current-summary}) and (\ref{energy current}) to 
$T$, $\mu$, $B$ by plugging in $M(T , \mu , B ) = \partial_B 
p_\text{thm} ( T , \mu , B )$ and use of the chain rule.

\subsection{Viscosities}

The viscosities have already been discussed at length in the
literature \cite{Bradlyn:2012,Jensen:2012}, but we rederive their Kubo
formulas in our language for completeness. Our treatment is
particularly close to that of Ref.~\cite{Jensen:2012}. Consider a rapid
metric perturbation $\delta h_{\mu \nu}$. Using the definition
(\ref{shear}) of the shear, we find
\begin{align}\label{stress variation}
	\delta T^{\mu \nu} &= ( \partial_T p \delta T + \partial_\mu p 
\delta \mu + \partial_B p \delta B - \zeta \delta \Theta ) g^{\mu \nu} 
\nonumber \\
	&\qquad - p \delta h^{\mu \nu} 
	- i \omega \eta {\Pi^{\mu \nu}}_{\lambda \rho} \delta 
h^{\lambda \rho}
	- i \omega \tilde  \eta  { { \tilde \Pi}^{\mu 
\nu}}_{\phantom{\mu \nu} \lambda \rho} \delta h^{\lambda \rho} .
\end{align}
$\delta T$ and $\delta \mu$ may of course be solved for using the 
linearized equations of motion but we will not need to do so here. 
${\Pi^{\mu \nu}}_{\lambda \rho}$ and ${\tilde \Pi^{\mu 
\nu}}_{\phantom{\mu \nu}\lambda \rho}$ are the even and odd symmetric 
traceless projectors
\begin{align}
	\Pi^{\mu \nu \lambda \rho}  = g^{\mu ( \lambda} g^{\rho ) \nu} 
- \frac{1}{2} g^{\mu \nu} g^{\lambda \rho}, \qquad
	\tilde \Pi^{\mu \nu \lambda \rho} = \frac{1}{2} \big( g^{\mu ( 
\lambda} \varepsilon^{\rho ) \nu} + g^{\nu ( \lambda} 
\varepsilon^{\rho ) \mu } \big) .
\end{align}
They are traceless in the first and second pairs of indices, have 
cross traces
\begin{align}
	{\Pi^{\mu \nu}}_{\mu \nu} = 2, \qquad
	{\tilde \Pi^{\mu \nu}}_{\phantom{\mu \nu}\mu \nu} = 0
\end{align}
and satisfy the algebra
\begin{align}
	{\Pi^{\mu \nu}}_{\alpha \beta} {\Pi^{\alpha \beta}}_{\lambda 
\rho} = {\Pi^{\mu \nu}}_{\lambda \rho} , \qquad
	{\Pi^{\mu \nu}}_{\alpha \beta} {\tilde \Pi^{\alpha 
\beta}}_{\phantom{\alpha \beta} \lambda \rho} = {\tilde \Pi^{\mu 
\nu}}_{\phantom{\alpha \beta} \lambda \rho} , \qquad
	{\tilde \Pi^{\mu \nu}}_{\phantom{\mu \nu} \alpha \beta} 
{\tilde \Pi^{\alpha \beta}}_{\phantom{\alpha \beta} \lambda \rho} = -  
{\Pi^{\mu \nu}}_{\lambda \rho} .
\end{align}

Using these identities and the symmetry properties $\Pi^{\mu \nu 
\lambda \rho} = \Pi^{\lambda \rho \mu \nu}$ and $\tilde \Pi^{\mu \nu 
\lambda \rho} = - \tilde \Pi^{\lambda \rho \mu \nu}$, it is then 
straightforward to verify that
\begin{subequations}
\begin{align}
	&\eta = - \lim_{\omega \rightarrow 0} \text{Im}\, 
\frac{\Pi_{\mu \nu \lambda \rho} G^{\mu \nu , \lambda \rho} (\omega) }
{2 \omega}\, , \\
	&\tilde \eta = - \lim_{\omega \rightarrow 0} \frac{\tilde 
\Pi_{\mu \nu \lambda \rho} G^{\mu \nu , \lambda \rho}(\omega)}{2 i 
\omega} \,.
\end{align}
\end{subequations}
Here and further, whenever we write $\lim_{\omega\to0}$, we assume
that spatial momentum is put to zero ($q=0$) before the limit is
taken.  Vice versa, when we write $\lim_{q\to0}$ we implicitly assume
that the frequency has been put to zero ($\omega=0$) befor the limit
is taken.

To get at the bulk viscosity, use $\delta \Theta = \frac{1}{2} i 
\omega g^{\mu \nu} \delta h_{\mu \nu}$ and
take the trace and imaginary part of (\ref{stress variation})
\begin{align}
	\zeta = - \lim_{\omega \rightarrow 0} \text{Im}\,\frac{G_{\mu 
\phantom{\mu , }\nu}^{\phantom \mu \mu , \phantom \nu \nu}(\omega)}{2 
\omega} \, .
\end{align}

\subsection{Thermal Conductivities}\label{sec:thermal conductivities}
Before deriving the remaining Kubo formulas, we would like to make 
some comments on the relation $\Sigma_T = - \frac{1}{T} \Sigma_G$ 
obtained from the entropy current analysis and rederive it from an 
alternative point of view that highlights the underlying physics. The 
Einstein relation identifies the conductivity and dissipation $\sigma 
= - \chi_\mu$ of any charged fluid 
where
\begin{align}
	j_i = \sigma E_i + \chi_\mu \partial_i \mu + \cdots .
\end{align}
These seemingly unrelated coefficients are connected by the following 
physical consideration. Apply a static but spatially varying electric 
potential $\delta A_0$. Charges will flow, but give the system time to 
relax and the current will again vanish. The chemical potential will 
adjust to match the profile of the electric potential $\delta \mu = 
\delta A_0$. Consistency then demands that $\sigma = - \chi_\mu$.

$\Sigma_T = - \frac{1}{T} \Sigma_G$ follows along similar lines. In 
the presence of the gravitational potential $-\Phi$, energy will flow 
from regions of large potential to small potential until equilibrium 
is reached. From appendix \ref{append:mag} we have for a static 
background
\begin{align}\label{def:temp}
	\beta = \int_c n , \qquad
	\mu = T \int_c A ,
\end{align}
where $c$ is the time circle passing through the fluid element under 
consideration. Now add a time independent perturbation $\delta n_\mu = 
- \delta \Phi n_\mu$ and the temperature and chemical potential adjust 
by 
\begin{align}
	\delta T = T \delta \Phi , \qquad
	\delta \mu = \mu \delta \Phi.
\end{align}
Altogether we have
\begin{align}
	\delta \varepsilon^\mu = ( \epsilon +T \partial_T \epsilon + 
\mu \partial_\mu \epsilon ) \delta \Phi v^\mu +  i q^\mu ( T \Sigma_T 
+ \Sigma_G )\delta \Phi + i ( T \tilde \Sigma_T + \mu \tilde 
\Sigma_\mu ) \varepsilon^{\mu \nu} q_\nu \delta \Phi .
\end{align}
We now impose consistency with the linearized equation of motion
\begin{align}
	\nabla_\mu \delta \varepsilon^\mu = - q^2 ( T \Sigma_T + 
\Sigma_G ) \delta \Phi = 0,
\end{align}
which gives the gravitational Einstein relation $\Sigma_T = - \frac{1}
{T} \Sigma_G$.

Luttinger first used the gravitational Einstein relation to obtain a 
Kubo formula for $\Sigma_T$ \cite{Luttinger:1964}. We perform a 
derivation in our language for completeness. Take a rapid transverse 
perturbation $\delta n_\mu$ such that $v^\mu \delta n_\mu = 0$,
\begin{align}
	\delta \varepsilon^\mu = ( \partial_T \epsilon \delta T + 
\partial_\mu \epsilon \delta \mu) v^\mu + i \omega \Sigma_G \delta 
n^\mu + i \omega \tilde \Sigma_G \varepsilon^{\mu \nu} \delta n_\nu ,
\end{align}
where we have used $\varepsilon^{\mu \nu} k_\nu = 0$. Upon application 
of projectors we have
\begin{align}
	&{P^\mu}_\lambda {P^\nu}_\rho G^{\lambda , \rho}_{\varepsilon 
\varepsilon} (\omega ) = i \omega ( \Sigma_G g^{\mu \nu} + \tilde 
\Sigma_G \varepsilon^{\mu \nu} ) \nonumber \\
	\implies \qquad
	\Sigma_T = -&\lim_{\omega \rightarrow 0} \text{Im}\, \frac{1}
{T}\frac{  g_{\mu \nu} G_{\varepsilon \varepsilon }^{\mu, \nu} (\omega 
) }{2 \omega}\,, 
	\qquad
	\tilde \Sigma_G = \lim_{\omega \rightarrow 0} 
\frac{\varepsilon_{\mu \nu} G^{\mu , \nu}_{\varepsilon \varepsilon} 
(\omega) }{2 i \omega} \,.
\end{align}

Finally, we derive a Kubo formula for the function $\tilde g$. Under a 
slow perturbation $\delta n_\mu = - \delta \Phi n_\mu$ we have
\begin{align}
	\delta T = T \delta \Phi , \qquad
	\delta \mu = \mu \delta \Phi , \qquad
	\delta G^\mu = i q^\mu \delta \Phi , \qquad
	\delta v^\mu = \delta \Phi v^\mu .
\end{align}
The energy current varies as
\begin{align}
	\delta \varepsilon^\mu = ( \epsilon + T \partial_T \epsilon + 
\mu \partial_\mu \epsilon ) \delta \Phi v^\mu
	+ i (T \tilde \Sigma_T + \mu \tilde \Sigma_\mu + \tilde 
\Sigma_G ) \varepsilon^{\mu \nu} q_\nu \delta \Phi .
\end{align}
Plugging in the explicit form of $\tilde \Sigma_T$ and $\tilde 
\Sigma_\mu$, we obtain
\begin{align}
	T \partial_T \tilde g + \mu \partial_\mu \tilde g = \lim_{q^2 
\rightarrow 0} \frac{\varepsilon_{\mu \nu}q^\mu  G^{\nu , 
\lambda}_{\varepsilon \varepsilon} ( q ) n_\lambda}{ i T^2 q^2} \, .
\end{align}

This only determines $\tilde g$ up to a function $f(\frac \mu  T)$. We 
can fix this ambiguity by response to a slow electric potential 
perturbation $\delta A_\mu = \delta a_0 n_\mu$. We then have
\begin{align}
	\delta T &= 0 , \qquad
	\delta \mu = \delta a_0 , \qquad
	\delta v^\mu = \frac{i}{B} \varepsilon^{\mu \nu} q_\nu \delta 
a_0 , \\
	\delta \varepsilon^\mu &= \left( \partial_\mu \epsilon \delta 
\mu \right) v^\mu + i \left( \tilde \Sigma_\mu + \frac{\epsilon}{B} 
\right) \varepsilon^{\mu \nu} q_\nu \delta a_0
\end{align}
giving
\begin{align}
	T^2 \partial_\mu \tilde g - \frac{p}{B}  = \lim_{q \rightarrow 
0} \frac{i \varepsilon_{\mu \nu} q^\mu G^{\nu , \lambda}_{\varepsilon 
j} (q) n_\lambda}{q^2} .
\end{align}
Recall here that derivatives are taken at constant $B$ rather than at 
constant $M$. The Kubo formulas for $\tilde \Sigma_G$ and $\tilde g$ 
completely determine the parity odd energy transport.

\section{Physical Interpretation}\label{sec:interpretation}

To compare with physical results, we first need deal with two issues. 
First, the covariant currents $\varepsilon^\mu$ and $T^{\mu \nu}$ have 
implicit dependence on $v^\mu$ that must be removed. This can be done 
by instead considering the noncovariant currents defined by
\begin{align}
	\delta S = \int d^3 x \sqrt{g} e^{-\Phi} \left( \frac{1}{2} 
T^{ij}_\text{nc} \delta g_{ij} + \varepsilon^0_\text{nc} \delta \Phi + 
\varepsilon^i_\text{nc} \delta C_i + j^\mu_\text{nc} \delta A_\mu 
\right) .
\end{align}
In Ref.~\cite{Geracie:2014} we demonstrate that these noncovariant currents 
are simply
\begin{align}\label{nc currents}
	j^\mu_\text{nc} = j^\mu ,\qquad
	\varepsilon^0_\text{nc} = e^{-\Phi} \varepsilon^0,
	\qquad
	\varepsilon^i_\text{nc} = e^{-\Phi} \varepsilon^i + 
{T^{ij}}v_j,
	\qquad
	T^{ij}_\text{nc} = T^{ij}
\end{align}
(these relations are greatly simplified by our use of the massless 
limit and selection of $s=1$).
The only change from the above is in the energy current, which we 
defer discussion of until later. Written out explicitly, we have
\begin{align}\label{noncovariant currents}
	j^0_\text{nc} &= e^\Phi n, \\
	j^i_\text{nc} &= e^\Phi \varepsilon^{ij} \left( \frac{n}{B} 
\Big(E_j - \d_j (e^{-\Phi} \mu )\Big) - \frac{s}{B}\d_j(e^{-\Phi}T) + 
\d_j (e^{-\Phi} M  )\right), \label{charge current1}\\
	T^{ij}_\text{nc} &= \left( p - \zeta \Theta \right) g^{ij} - 
\eta \sigma^{ij} - \tilde \eta \tilde \sigma^{ij} . 
\end{align}

The second issue is that to perform the LLL projection we have taken 
the $g$-factor to be $g=2$ (and the spin to be $s=1$ though this is 
not essential). To compare to standard expressions used in literature
we need to transform back to the values commonly assumed, $g=s=0$.
The result turns out to be rather 
trivial, in the end giving us back (\ref{noncovariant currents}) with 
shifted transport coefficients, but it is worth demonstrating how this 
comes about. In the process we find simple formulas that demonstrate 
how to recover the physical transport coefficients from those 
calculated in the massless limit.
The general procedure for how to do this is explained in 
Ref.~\cite{Geracie:2014} and we merely outline the results here.

Note that to simplify the resulting formulas we assume that $E_i$ is 
$\mathcal O (1)$ in derivatives. In the above, the electric field was 
potentially large; however, since it's variations are assumed to be 
small, a frame where $E_i$ is small everywhere may always be obtained. 
In such a frame a large number of terms are higher order and 
neglected. Indeed we have already used this in (\ref{nc currents}) to 
neglect terms that involve the mass which we are otherwise restoring.

The $g=s=0$ currents are then
\begin{subequations}\label{currents}
\begin{align}
	j^0_\text{nc} &= e^\Phi n, \\
	j^i_\text{nc} &= e^\Phi \varepsilon^{ij} \left( \frac{n}{B} 
\Big(E_j - \d_j \Big(e^{-\Phi}\big( \mu+ \frac{B}{2m} \big) \Big)\Big) 
- \frac{s}{B}\d_j(e^{-\Phi}T) + \d_j \Big(e^{-\Phi}\big( M - \frac{n}
{2m} \big) \Big)\right), \label{currents2} \\
	T^{ij}_\text{nc} &= \left( p + \frac{Bn}{2m} - \zeta \Theta 
\right) g^{ij} - \eta \sigma^{ij} - \left( \tilde \eta + \frac{1}{2} n 
\right) \tilde \sigma^{ij} . \label{currents3}
\end{align}
\end{subequations}
This has a simple interpretation and with a little physical insight we 
could have guessed the form given here. Recall from 
Ref.~\cite{Geracie:2014} that redefining $g$ involves a shift to the 
electric potential
\begin{align}
	A_0^{g=2} = A_0 - \frac{1}{2m} e^{-\Phi} B
\end{align}
This also shifts the ground state energy of the system and so the 
chemical potential changes
\begin{align}
	\mu^{g=2} = \mu - \frac{B}{2m} .
\end{align}
$\mu^{g = 2}$ is the chemical potential that appears in 
(\ref{currents2}).

Similarly, setting $g=2$ alters the intrinsic magnetic moment of the 
 fluid: each particle carries an excess magnetic dipole moment of 
$\frac{1}{2m}$. The $g=2$ and physical magnetizations are then related 
by
\begin{align}
	M^{g=2} = M + \frac{n}{2m}\,,
\end{align}
accounting for the final term in (\ref{currents2}) and the shift to 
the internal pressure in (\ref{currents3}) since $p = p_\text{thm} - 
MB$.
Finally, setting $s=1$ overestimates the intrinsic angular momentum 
per particle by $1$, giving the observed shift in the Hall 
viscosity. In the end, the constitutive relations simply revert to the 
form (\ref{noncovariant currents}) where we are using the $g=s=0$ 
values of $\mu$, $M$, and $\tilde \eta$.

The non-covariant currents then satisfy equations of motion 
\cite{Geracie:2014}
\begin{subequations}\label{noncovariant ward}
\begin{align}
	\frac{1}{\sqrt{g}} e^\Phi \partial_0 ( \sqrt{g} e^{-\Phi} 
j^0_\text{nc} ) + e^\Phi \nabla_i ( e^{-\Phi} j^i_\text{nc} ) &= 0, \\
	\frac{1}{\sqrt{g}}\partial_0 \big( \sqrt{g} 
\varepsilon^0_{\text{nc}} \big) + e^\Phi \nabla_i ( e^{-\Phi} 
\varepsilon^i_{\text{nc}} )&= E_i j^i_\text{nc} - \frac{1}{2} 
T^{ij}_{\text{nc}} \dot g_{ij} , \\
	\frac{e^\Phi}{\sqrt{g}} \partial_0 \big( \sqrt{g} \,
m j^i_\text{nc} 
\big)  + e^\Phi \nabla_j ( e^{-\Phi} 
{T_{\text{nc}i}}^j ) 
	&= j^0_\text{nc} E_i + \varepsilon_{ij} j^j_\text{nc} B +  
\varepsilon^0_{\text{nc}} \nabla_i \Phi .
\end{align}
\end{subequations}

\subsection{The Charge Current}
Now consider the current response to the electric field $\mathbf E$, 
the
gravitational field $\mathbf G = \bm \nabla \Phi$ and gradients of 
$T$, $\mu$ and $B$
\begin{align}
  \mathbf j_{\rm nc} &= \left(\sigma_H \mathbf E + \sigma_H^T\bm\nabla 
T  + \sigma_H^\mu \bm\nabla\mu + \sigma_H^B \bm \nabla B
    + \sigma_H^G \bm G \right)\times \mathbf{\hat z} .
\end{align}
We find a Hall conductance
\begin{equation}
\sigma_H = e^\Phi \frac n B .
\end{equation}
This equation is can be obtained trivially by going to the coordinate
system moving with the velocity $(\mathbf E\times\mathbf{\hat z})/B$,
in which the electric field vanishes. 

The Hall diffusivity is
\begin{equation}
  \sigma_H^\mu = \left(\frac{\d M}{\d\mu}\right)_{T,B} - \frac nB 
\end{equation}
which using Maxwell's relations can be written as
\begin{equation}\label{diffusivity}
  \sigma_H^\mu = \left(\frac{\d n}{\d B}\right)_{T,\mu} - \frac nB  .
\end{equation}
From this equation it is easy to argue the existence of Hall plateaus 
when the chemical potential lies in a gap. In the $T \rightarrow 0$ 
limit, small variations in the chemical potential cannot induce 
electron transport and so $\sigma^\mu_H = 0$. Equation 
(\ref{diffusivity}) then immediately implies
\begin{align}\label{Streda}
	n = \nu B
	\qquad \implies \qquad
	\sigma_H = \nu,
\end{align}
where we have taken $\Phi = 0$ and $\nu$ is some constant (which we of 
course know to be the filling fraction). (We are working in units 
where $e = \hbar = 1$.)

Similarly we also find
\begin{equation}
	\sigma_H^T = \left( \frac{\partial s}{\partial B} \right)_{T, 
\mu}- \frac{s}{B}\,, \qquad
	\sigma_H^B = \left( \frac{\partial M}{ \partial B } \right)_{T 
, \mu}, \qquad
	\sigma_H^G = \frac{\epsilon + p}{B}\,,
\end{equation}
so in particular $\sigma^B_H$ is simply the magnetic susceptibility.

\subsection{The Energy Current}

We now turn to energy transport. A redefinition
\begin{align}
	\tilde \Sigma_G &\equiv \frac{\mu}{B} \left( Ts + \mu n 
\right) + T^2 c_{RL} - 2 M_E , \qquad
	\tilde g \equiv \frac{M_E}{T^2}  \,,
\end{align}
will aid in the physical interpretation of the formulas that follow. 
Including the $T^{ij}v_j$ shift to the covariant energy current we 
have
\begin{align}\label{energy current1}
	\varepsilon^0_\text{nc} &= \epsilon \nonumber \\
	\varepsilon^i_\text{nc} &= \varepsilon^{ij} \bigg(  
\frac{\epsilon + p}{B} \left( E_j - \d_j (e^{-\Phi}  \mu  ) \right) - 
M \d_j (e^{-\Phi}  \mu ) 
	-  T c_{RL} \partial_j ( e^{-\Phi} T ) 
	+ e^\Phi \partial_j ( e^{-2 \Phi} M_E  )   \bigg) \nonumber \\
	&+ \Sigma_T \partial^i \left( e^{-\Phi} T \right) .
\end{align}
The $g=0$ values of the energy density $\epsilon$, energy 
magnetization $M_E$, and Righi-Leduc coefficient $c_{RL}$ that are 
used in this formula are related to the $g=2$ values by
\begin{align}
	\epsilon = \epsilon^{g=2} + \frac{Bn}{2m},
	\qquad
	M_E = M^{g=2}_E + \frac{M^{g=2}B}{2m},
	\qquad
	c_{RL} = c^{g=2}_{RL} + \frac{s/T}{2m}.
\end{align}

Defining thermal conductivities
\begin{align}
  \boldsymbol \varepsilon_{\rm nc} &= \kappa \bm \nabla T + 
\kappa^\Phi \bm G+ \left(\kappa_H \mathbf E + 
\kappa_H^T\bm\nabla T + \kappa_H^\mu \bm\nabla\mu
    + \kappa_H^B \bm \nabla B+ \kappa_H^G \bm G \right)\times 
\mathbf{\hat z}
\end{align}
we have
\begin{align}
	\kappa &= \Sigma_T \qquad
	\kappa^\Phi = - T \Sigma_T \qquad
	\kappa_H = \frac{\epsilon + p}{B} \nonumber \\
	\kappa_H^T &= e^{-\Phi} \left( \partial_T M_E - T c_{RL} 
\right)  \qquad
	\kappa_H^\mu = e^{-\Phi} \left( \partial_\mu M_E - \frac{Ts + 
\mu n}{B} \right) \nonumber \\
	\kappa_H^B &=  e^{-\Phi} \partial_B M_E \qquad 
	\kappa_H^G = e^{-\Phi} \left( T^2 c_{RL} - 2 M_E + \frac{\mu }
{B} ( Ts + \mu n ) \right) .
\end{align}

\subsection{St\v reda Formulas}
One notable feature about the formulas (\ref{charge current1}) and 
(\ref{energy current1}) is the charge and energy currents that persist 
in thermal equilibrium. We now turn to these, deriving a set of St\v 
reda-like formulas for two dimensional fluids. First note from the 
definitions of the temperature and chemical potential in appendix 
\ref{append:mag} that in thermal equilibrium we have $\partial_i \mu = 
e^\Phi E_i + \mu G_i$ and $\partial_i T = T G_i$ where $G_i = 
\partial_i \Phi$ is the gravitational field exerted by the potential 
$-\Phi$. The equilibrium currents are then
\begin{align}
	j^i_\text{nc} = \varepsilon^{ij} e^\Phi \d_j ( e^{-\Phi} M ) 
\qquad
	\varepsilon^i_\text{nc} = \varepsilon^{ij} \left( - M \d_j ( 
e^{-\Phi} \mu ) + e^\Phi\d_j (e^{-2\Phi} M_E ) \right).
\end{align}
Expressing these in terms of the externally applied fields $E_i$, $B$ 
and $G_i$ we have
\begin{align}\label{persistent}
	j^i_\text{nc} &= \varepsilon^{ij} \big( e^\Phi \d_\mu M E_j + 
\d_B M \d_j B + (T \partial_T M + \mu \d_\mu M - M )G_j \big) 
\nonumber \\
	\varepsilon^i_\text{nc} &= \varepsilon^{ij} \big( ( \d_\mu M_E 
- M ) E_j + e^{-\Phi} \d_B M_E \d_j B + e^{-\Phi}(T \d_T M_E + \d_\mu 
M_E - 2 M_E ) G_j \big) .
\end{align}

Defining equilibrium responses by
\begin{align}
	j^i_\text{nc} &=  \varepsilon^{ij} \big( \sigma^{\text{eq}}_H 
E_j  +  \sigma^{B\text{eq}}_H \d_j B + \sigma^{G\text{eq}}_H G_j \big),
\end{align}
and using some Maxwell relations, we have
\begin{align}
	\sigma^\text{eq}_H &= \left( \frac{\d n}{\d B} \right)_{T,\mu} ,
\qquad
	\sigma^{B\text{eq}}_H = \left( \frac{\d M}{\d B} 
\right)_{T,\mu} , \nonumber \\
	\sigma^{G\text{eq}}_H &= T \left( \frac{\d s}{\d B} 
\right)_{T,\mu} + \mu \left( \frac{\d n}{\d B}\right)_{T , \mu} - M .
\end{align}
where we have set $\Phi = 0$ in these formulas. The first is the 
well-known St\v reda formula \cite{Streda:1982}. The following two are St\v 
reda-like formulas for currents induced by inhomogeneities in $B$ and 
external gravitational forces. 

Similarly working with the energy current
\begin{align}
	\varepsilon^i_\text{nc} &=  \varepsilon^{ij} \big( 
\kappa^{\text{eq}}_H E_j  +  \kappa^{B\text{eq}}_H \d_j B + 
\kappa^{G\text{eq}}_H G_j \big) ,
\end{align}
we find that all equilibrium currents are determined by the 
magnetization $M$ and energy magnetization $M_E$.
\begin{align}
	\kappa^\text{eq}_H &= \left( \frac{\d M_E}{\d \mu} 
\right)_{T,B} - M ,\qquad
	\kappa^{B\text{eq}}_H = \left( \frac{\d M_E}{\d B} 
\right)_{T,\mu} , \nonumber \\
	\kappa^{G\text{eq}}_H &= T\left( \frac{\d M_E}{\d T} 
\right)_{\mu , B}  + \mu \left( \frac{\d M_E}{\d \mu} \right)_{T , B} 
- 2 M_E .
\end{align}
A similar collection of St\v reda formulas was recently presented in 
Ref.~\cite{Gromov:2014}; however they do not agree with ours.

Finally, we note that these St\v reda formulas in no way depend on the 
LLL projection that has been implicit throughout this paper. Indeed, 
they follow only from knowledge of the equilibrium persistent currents 
(\ref{persistent}), which may be derived on entirely general grounds. 
Assuming a static background and dynamic equilibrium, current 
conservation implies that $j^i_\text{nc}$ is the curl of a function 
which we identify as the magnetization density
\begin{align}
	\nabla_i  j^i = 0
	\qquad \implies \qquad
	j^i_\text{nc} = \varepsilon^{ij} \d_j  M ,
\end{align}
where we have taken $\Phi = 0$. We can then retrieve the energy 
current by demanding a static energy density $\d_0 
\varepsilon^0_\text{nc} = 0$
\begin{align}
	\nabla_i  \varepsilon^i_\text{nc} = E_i j^i_\text{nc}
	\qquad \implies \qquad
	\nabla_i  \varepsilon^i_\text{nc} = - \nabla_i (M 
\varepsilon^{ij} E_j ),
\end{align}
where we have used that $\varepsilon^{ij} \nabla_i E_j = - \dot B = 
0$. We then similarly obtain
\begin{align}
	\varepsilon^i_\text{nc} = \varepsilon^{ij} \big( - M E_j + 
\d_j M_E \big) 
\end{align}
for some function $M_E$, reproducing (\ref{persistent}). 
This equilibrium current is also found in Ref.~\cite{Cooper:1997}.
Were we to 
carry this out for nonzero $\Phi$, we would again obtain the correct 
result, but the normalization of $M$ and $M_E$ with factors of $e^{-
\Phi}$ has to be determined from other considerations.

\subsection{Noncovariant Kubo Formulas}\label{sec:kubo nc}
For the reader's convenience, we restate here the Kubo formulas found 
above in terms of the energy magnetization $M_E$ and thermal Hall 
coefficient $c_{RL}$ without the use of the Newton-Cartan formalism. 
They are expressed in terms of two-point correlators of the non-
covariant currents 
\begin{align}
	\mathcal G^{\mu , \nu}_{\varepsilon \varepsilon} (x) &= \left 
\langle 
\frac{\delta \varepsilon^\mu_\text{nc} (x)}{ \delta n_\nu ( 0 )} 
\right \rangle 
- i \theta ( x^0 ) \left \langle [ \varepsilon^\mu_\text{nc} ( x) , 
\varepsilon^\nu_\text{nc} ( 0 )] \right \rangle , \nonumber  \\
	\mathcal G^{\mu , \nu}_{\varepsilon j} (x) &= \left \langle 
\frac{\delta \varepsilon^\mu_\text{nc} (x)}{ \delta A_\nu ( 0 )} 
\right \rangle 
+ i \theta ( x^0 ) \left \langle [ \varepsilon^\mu_\text{nc} ( x) , 
j^\nu_\text{nc} ( 0 )] 
\right \rangle , \nonumber\\
	\mathcal G^{i j , k l}(x) &= \left \langle \frac{\delta 
T^{i j}_\text{nc} (x)}{ \delta g_{k l} ( 0 )} \right \rangle + 
\frac{i}{2} \theta ( x^0 ) \left \langle [ T^{i j}_\text{nc} ( x) , 
T^{k l}_\text{nc} ( 0 )] \right \rangle  .
\end{align}
These are
\begin{equation}
	\eta = - \lim_{\omega \rightarrow 0} \text{Im} 
\frac{\Pi_{ijkl} \mathcal G^{ij,kl} ( \omega )}{2 \omega} ,\qquad
	\tilde \eta = - \lim_{\omega \rightarrow 0} \frac{\tilde 
\Pi_{ijkl} \mathcal G^{ij , kl}(\omega)}{2 i \omega},\qquad
	\zeta = - \lim_{\omega \rightarrow 0} 
\text{Im}\,\frac{\mathcal G_{i \phantom{i , }j}^{\phantom i i , 
\phantom j j}(\omega)}{2 \omega} ,\nonumber 
\end{equation}
\begin{align}
	&\partial_\mu M_E -M = \lim_{q \rightarrow 0} \frac{i 
\varepsilon_{ij} q^i \mathcal G^{j,0}_{\varepsilon j} (q)}{q^2},
	&&T \partial_T M_E + \mu \partial_\mu M_E - 2 M_E = - \lim_{q 
\rightarrow 0} \frac{i \epsilon_{ij} q^i \mathcal G^{j,0}_{\varepsilon 
\varepsilon} (q)}{q^2}, \nonumber \\
	&\Sigma_T = \frac{i}{T} \lim_{\omega \rightarrow 0} 
\frac{\delta_{ij} \mathcal G^{ij}_{\varepsilon \varepsilon} ( \omega 
)}{2 \omega},
	&&T^2 c_{RL} + \frac{\mu}{B} (Ts + \mu n ) - 2 M_E = 
\lim_{\omega \rightarrow 0} \frac{\epsilon_{ij} \mathcal 
G^{ij}_{\varepsilon \varepsilon} ( \omega )}{2 i \omega},
\end{align}
the correlators being evaluated on the trivial background $g_{ij} = 
\delta_{ij}$, $\Phi = 0$, $E_i = 0$ and $\partial_i B = 0$.

\section{Conclusion}
\label{sec:concl}

The proper coordinate invariant description of non-relativistic
physics is that of a Newton-Cartan geometry, which naturally includes
a source $n_\mu$ for the energy current in addition to those present
for the stress and charge current. As discussed in recent work, with
some care, diffeomorphism covariant currents may then be defined and
1-point Ward identities follow naturally as in the nonrelativistic
case.

In a fluid dynamical description, the Ward identities become equations 
of motion once constitutive relations have been supplied. We have 
given the most general constitutive relations consistent with 
diffeomorphism covariance and derived their Kubo formulas. We argue 
that a fractional quantum Hall fluid is distinguished as being a {\it 
force-free} fluid in $2+1$ dimensions. The force-free condition 
immediately gives powerful constraints on fractional quantum Hall 
transport, determining all charge transport in terms of 
thermodynamics.

A straightforward entropy current analysis was then performed. The 
expected restrictions on the signs of parity even viscosities and 
thermal conductivity are obtained, in addition to new constraints on 
the transverse energy response. These four coefficients are not 
independent but are instead determined by two free functions of $T$, 
$\mu$ and $M$: the thermal Hall conductivity and energy magnetization. 
The derived constitutive relations imply a set of formulas for the 
equilibrium response that generalize the well-known St\v reda formula. 
These new formulas characterize the system's response to Newtonian 
gravitational fields and inhomogeneous magnetic backgrounds.

It is our hope that the approach outlined here to non-relativistic 
fluids finds further use. In this approach spacetime coordinate 
invariance 
is automatic, just as in the standard 
treatment of relativistic fluids and computations are streamlined. 
Here we brought our formalism to bear on FQH fluids, but it is 
sufficiently general to treat arbitrary fluids in any dimension.

\acknowledgments

We would like to thank Alexander Abanov, Andrei Gromov, Kristan Jensen
and and Paul Wiegmann for discussions.  This work is supported, in part,
by the US DOE grant No. DE-FG02-13ER41958, and the ARO-MURI
63834-PH-MUR grant, and a Simon Investigator grant from the Simons
Foundation.

\begin{appendices}

\section{Magneto-Thermodynamics}\label{append:mag}
We make a few comments on thermodynamics in curved backgrounds and 
nonzero magnetic fields to motivate the identifications of 
$\epsilon$, $n$ and $p$ in (\ref{equilibrium current 1}) and 
(\ref{equilibrium current 2}) as the thermodynamic energy density, 
particle density and internal pressure, particularly the magnetization 
contribution whose presence is not obvious. These issues are discussed 
at length in Ref.~\cite{Hartnoll:2007} in the case of free field theory, 
but we take a more general view, assuming only local thermodynamic 
equilibrium and a local free energy. This is the zeroth order part of 
an analysis along the lines of that found in Ref.~\cite{Kaminski:2012}, 
from which we differ only in so far as our treatment is non-
relativistic and includes a background magnetic field that gives rise 
to the magnetization currents in question.

Consider the thermal partition function $Z$ associated to some 
microscopic quantum field theory and it's corresponding effective 
action
\begin{align}
	W = - \ln Z.
\end{align}
We assume nothing about the detailed dynamics other than the existence 
of a gap so that $W$ is a local function of $n_\mu$, $g^{\mu \nu}$ and 
$A_\mu$.
Specialize to a time independent but curved background geometry and 
gauge field whose spatial variations are small. We may then assume 
after some time local thermodynamic equilibrium is reached and each 
fluid element is characterized entirely by a temperature and a 
chemical potential
\begin{align}
	\beta = \int_c n , \qquad
	\frac{\mu}{T} = \int_c A ,
\end{align}
where $c$ is the time circle passing through that element. Note that 
$T$ and $\mu$ may depend on space.

To zeroth order in derivatives, we then have
\begin{align}
	W = \int\! d^3 x\, \sqrt{g}\, e^{-\Phi} p_\text{thm} ( T , \mu 
, B ),
\end{align}
$B$ being the only other covariant scalar that may be constructed at
zeroth order. The detailed form of $p_\text{thm}$ will depend on the
microscopic physics but will not be needed here. Had we assumed
spatial homogeneity, this would merely be the elementary relation
$\Omega = p_\text{thm} V$ that connects thermodynamics with 
statistical physics
($\Omega = T W $ is the grand potential). Thus 
$p_\text{thm}$ is the grand potential density which, in the absence
of the magnetic field, would coincide with the pressure that appears 
in 
the stress.  
We define local
energy, entropy, particle and magnetization densities by
\begin{align}
	dp_\text{thm} = s dT + n d \mu + M dB
	\qquad \text{and} \qquad
	\epsilon + p_\text{thm} = Ts + \mu n,
\end{align}
which are merely the fundamental thermodynamic relations 
(\ref{thermodynamics}).

It's now a simple matter to calculate the equilibrium $j^0$, 
$\varepsilon^0$ and $T^{ij}$. To clarify the $\Phi$ dependence, 
parameterize the time circle by some interval $x^0 \in (0, \frac 1 
{T_0})$. We then have
\begin{align}
	T = e^\Phi T_0, \qquad \mu = e^\Phi A_0 .
\end{align}
Varying $A_0$, $\Phi$ and $g^{ij}$ we find
\begin{align}
	j^0 = e^\Phi n, \qquad
	\varepsilon^0 = e^\Phi \epsilon, \qquad
	T^{ij} = ( p_\text{thm} - M B) g^{ij}
\end{align}
The magnetization contribution to the internal pressure arises due to 
the magnetic flux density's metric dependence $B = \frac{1}{\sqrt{g}} 
( \partial_1 A_2 - \partial_2 A_1)$.

\section{Weyl Invariance}\label{append:weyl}

In special cases the theory may exhibit Weyl invariance.  This
happens, for example, when the interaction is a purely contact
interaction~\cite{Geracie:2014}.  In this case the functional form of
the transport coefficients considered above will be constrained. We
derive these constraints in this appendix.

A Weyl invariant theory is unchanged under the transformation
\begin{align}
	g'_{ij} = e^{-2 \alpha} g_{ij}, \qquad
	\Phi' = \Phi + 2 \alpha .
\end{align}
Since we have $S[g_{ij} ,\Phi] = S[g'_{ij} ,\Phi']$, varying the 
action we find 
\begin{align}\label{scaling}
	T'^{i j}_\text{nc} = e^{6 \alpha} T^{\mu \nu}_\text{nc}, \qquad
	\varepsilon'^\mu_\text{nc} = e^{4 \alpha} 
\varepsilon^\mu_\text{nc},
\end{align}
for a Weyl invariant theory. From the equilibrium definitions of the 
thermodynamic variables we also find
\begin{align}
	T' = e^{2 \alpha} T, \qquad
	\mu' = e^{2 \alpha} \mu, \qquad
	B' = e^{2 \alpha} B .
\end{align}

Let's first turn to the stress tensor
\begin{align}
	T^{ij}_\text{nc} = (p - \zeta \Theta ) g^{ij} - \eta 
\sigma^{ij} - \tilde \eta \tilde \sigma^{ij} .
\end{align}
One may show from their definitions that the expansion and shear 
tensors transform as
\begin{align}
	\Theta' = e^{2 \alpha} \left( \Theta - 2 v^\mu \nabla_\mu 
\alpha \right), \qquad
	\sigma'^{ij} = e^{4 \alpha} \sigma^{ij}, \qquad
	\tilde \sigma'^{ij} = e^{4 \alpha} \tilde \sigma^{ij} .
\end{align}
To satisfy the scaling rule (\ref{scaling}) the bulk viscosity must 
vanish: $\zeta = 0$~\cite{Son:2007}. Furthermore, the equation of state 
and viscosities must be homogeneous functions of the thermodynamic 
variables
\begin{align}
	p_\text{thm} ( \lambda T , \lambda \mu , \lambda B ) &= 
\lambda^2 p_\text{thm} ( T , \mu , B), \nonumber \\
	\eta ( \lambda T , \lambda \mu , \lambda B ) &= \lambda \eta ( 
T , \mu , B ), \nonumber \\
	\tilde \eta ( \lambda T , \lambda \mu , \lambda B ) &= \lambda 
\tilde  \eta ( T , \mu , B ) .
\end{align}

Similar restrictions arise for the energy current without 
complication. The thermal conductivity, thermal Hall conductivity, and 
energy magnetization are also homogeneous functions,
\begin{align}
	\Sigma_T ( \lambda T , \lambda \mu , \lambda B ) &= \lambda 
\Sigma_T ( T , \mu , B ), \nonumber \\
	c_{RL} ( \lambda T , \lambda \mu , \lambda B ) &= c_{RL} ( T , 
\mu , B ), \nonumber \\
	M_E ( \lambda T , \lambda \mu , \lambda B ) &= \lambda^2 M_E ( 
T , \mu , B) .
\end{align}

\end{appendices}

\end{document}